\documentclass[a4paper,11pt]{article}
\pdfoutput=1 

\usepackage{jheppub} 

\usepackage[T1]{fontenc} 

\usepackage{amssymb}
\usepackage{amsthm}
\usepackage{xspace}
\usepackage{bookmark}
\usepackage{tabstackengine}
\usepackage{todonotes}
\usepackage{subfig}
\allowdisplaybreaks

\def\ep  {\ensuremath{\epsilon}}

\def\OO {\ensuremath{O}}
\def\s {\bar{s}}
\def\II {\text{II}}

\title{Three-loop photon spectral density in QED.}

\author[a,b,c]{A.I. Onishchenko}

\affiliation[a]{Bogoliubov Laboratory of Theoretical Physics, Joint
	Institute for Nuclear Research,\\ Dubna, Russia}
\affiliation[b]{Skobeltsyn Institute of Nuclear Physics, Moscow State University\\ Moscow, Russia}
\affiliation[c]{Theory department, Budker Institute of Nuclear Physics,\\ Novosibirsk, Russia}

\emailAdd{onish@theor.jinr.ru}

\usepackage{blindtext}

\abstract{
We calculate three-loop photon spectral density in QED with $N$ different species of electrons. The obtained results were expressed in terms of iterated integrals, which are either  reduce to Goncharov's polylogarithms or can be written in terms of one-fold integrals of harmonic polylogarithms and complete elliptic integrals. In addition we provide threshold and high-energy asymptotics of the calculated spectral density. It is shown, that the use of the obtained spectral density correctly reproduces separately calculated moments of corresponding photon polarization operator.  	 
}

\begin{document} 
\maketitle
\flushbottom
\section{Introduction}

Photon self-energy operator $\Pi (s)$, defined as an extra factor $(1+\Pi (s))$ in the denominator of renormalized photon propagator, is one of the several fundamental quantities arising in a study of quantum electrodynamics (QED). At present we have exact results for one- and two-loop contributions \cite{berestetskii1982quantum,kallen1955k}. However, starting from three loops there are only approximate results \cite{Baikov:1995ui,kinoshita1999sixth,kinoshita1999accuracy,kinoshita1983parametric}. For example, the derivation of Baikov and Broadhurst \cite{Baikov:1995ui} employs a simple Pad\`e approximation  for three loop contribution using a few terms of the asymptotic expansions near 3 special points: $s=0,4m^2,\infty$ ($m$ is the electron mass).

In this paper we present exact results for three-loop photon spectral density. The latter is defined as a discontinuity of photon polarization operator. The obtained exact and asymptotic expressions for spectral density are then used for the calculation of the moments of photon polarization operator. In addition to this we provide results for first hundred terms of generalized Frobenius power series expansion of photon polarization operator at $s = 0$. The knowledge of exact photon spectral density already allows us to check the importance of the missed threshold ($s=16m^2$) in the reconstruction analysis of three-loop photon polarization operator performed by Baikov and Broadhurst. Still, we decided to explore this question in more detail later when we will have exact and asymptotic results for the polarization operator itself.

\begin{figure}
	\includegraphics[width=1\textwidth]{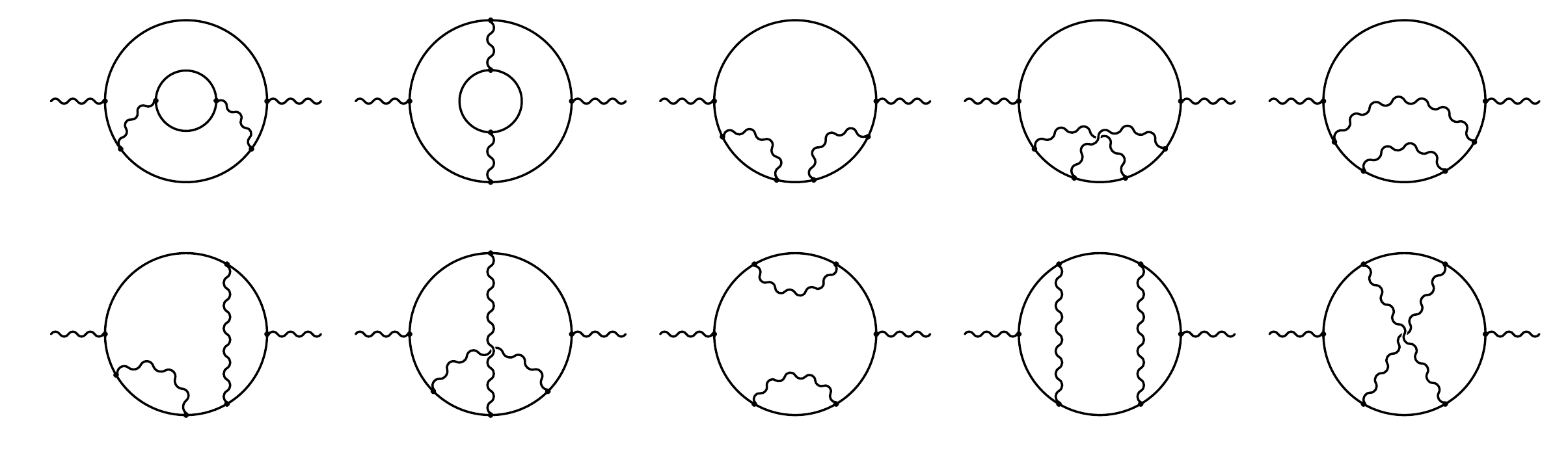}
	\caption{Diagrams contributing to $3$-loop photon self energy in QED.}
	\label{fig:diagrams}
\end{figure}

\section{Spectral density calculation}

To calculate three-loop QED photon spectral density $\rho (s)$  we followed standard procedure: 
\begin{itemize}
\item generation of Feynman diagrams\footnote{For this purpose we used FeynArts package \cite{FeynArts}.} for photon self-energy (see Fig. \ref{fig:diagrams})
\item application of projector to extract photon polarization operator $\Pi (s)$ and subsequent mapping of scalar integrals to the minimal set of prototype integrals
\item  IBP reduction \cite{ChetTka1981,Tkachov1981} of prototype integrals to the set of master integrals and application of bipartite cuts with Cutkosky rules for the latter
\item substitution of expressions for cut master integrals\footnote{The calculation of one and two loop cut master integrals is similar to  \cite{3loop-masters}.} \cite{Lee:2019wwn, 3loop-masters} and renormalization.
\end{itemize}
This way, considering QED with $N$ electron flavors\footnote{The masses of all electrons are assumed to be equal.} and using on-shell renormalization scheme\footnote{See Appendix \ref{renorm-constants} for results of calculation of required renormalization constants.} we get
\begin{equation}
\rho (s) = \rho^{(1)} (s) \left(\frac{\alpha}{4\pi}\right) + 
\rho^{(2)} (s) \left(\frac{\alpha}{4\pi}\right)^2 + \rho^{(3)} (s) \left(\frac{\alpha}{4\pi}\right)^3 + \ldots\, ,
\end{equation}
where ($\s = \frac{s}{m^2}$, $\beta = \sqrt{1-\frac{4}{\s}}$):
\begin{align}
\rho^{(1)}(s) =& \frac{4 N\pi (2+\s)\beta }{3\s}\, , \\
\rho^{(2)}(s) =& \frac{16 N\pi}{3}\Big\{
\beta^2 \left(2 \II_{l_1,l_2} + 2\II_{l_2,l_1}+\II_{l_1,l_0}+\II_{l_0,l_1}\right)
-\frac{\beta (2+\s)}{\s}\left( \II_{l_0}+2\II_{l_2}\right) \nonumber \\
& - \frac{7+8\log 2 + \s (2-(3+\log 4)\s)}{\s^2}\II_{l_1} + \frac{4\pi^2 (\s^2-4)-\beta\s (8 (2+\s)\log 2 - 3 (6+\s))}{4\s^2}
\Big\}\, .
\end{align}
The three-loop contribution $\rho^{(3)}(s)$ can be naturally separated into two pieces corresponding to $2m$ and $4m$ cuts, so that we have
\begin{equation}
\rho^{(3)}(s) = \rho^{(3)}_{2m}(s) + \theta (s - 16 m^2)\rho^{(3)}_{4m}(s)\, ,
\end{equation}
where
\begin{align}
\rho^{(3)}_{2m}(s) =& 
\frac{16\pi N^2}{3}\Big\{-\frac{4 c_{25}}{27}+\frac{2}{27} c_{30} \text{II}_{l_1}+\frac{4}{9} c_{27}
\text{II}_{l_1,l_1}-\frac{2}{3} c_9 \text{II}_{l_1,l_1,l_1}\Big\}
+\frac{16\pi N}{3}\Big\{\frac{c_{45}}{12}+\frac{1}{2} c_{36} \text{II}_{l_0}\nonumber \\ & -\frac{1}{24} c_{46}
\text{II}_{l_1}-c_{35} \text{II}_{l_2}+\frac{1}{6} c_{37} \text{II}_{l_0,l_1}-\frac{1}{4}
c_{39} \text{II}_{l_1,l_0}-c_{34} \text{II}_{l_1,l_1}+\frac{1}{2} c_{38}
\text{II}_{l_1,l_2}+4 c_{43} \text{II}_{l_2,l_1}
\nonumber \\ &
+8 c_2 \left(\text{II}_{l_0,l_0}+2
\left(\text{II}_{l_0,l_2}+\text{II}_{l_2,l_0}+2 \text{II}_{l_2,l_2}\right)\right)+2 c_{15}
\text{II}_{l_0,l_0,l_1}-2 c_5 \text{II}_{l_0,l_1,l_1}
\nonumber \\ &
+4 c_3 \left(\text{II}_{l_0,l_1,l_0}+2
\text{II}_{l_0,l_1,l_2}\right)-2 c_{42} \text{II}_{l_1,l_0,l_1}+c_{44}
\text{II}_{l_1,l_1,l_1}+4 c_{40} \left(\text{II}_{l_1,l_1,l_0}+2
\text{II}_{l_1,l_1,l_2}\right)\nonumber \\ & -4 c_{41} \text{II}_{l_1,l_2,l_1}-4 c_6
\text{II}_{l_2,l_1,l_1}-8 c_{12} \left(\text{II}_{l_1,l_0,l_0}+2 \text{II}_{l_1,l_0,l_2}+2
\text{II}_{l_1,l_2,l_0}+4
\text{II}_{l_1,l_2,l_2} \right.
\nonumber \\ &
\left. +\text{II}_{l_2,l_0,l_1}
-\text{II}_{l_2,l_1,l_0}-2
\text{II}_{l_2,l_1,l_2}-2 \text{II}_{l_2,l_2,l_1}\right)
-8 c_{17}
\left(\text{II}_{l_0,l_2,l_1}+\text{II}_{l_0,l_1,l_1,l_1}\right)
\nonumber \\ &
+c_{21} \text{II}_{l_1,l_1,l_0,l_1}
+c_{11} \left(3 \left(2 \log (2)
\text{II}_{l_4,l_1}+\text{II}_{l_4,l_1,l_0}+2
\text{II}_{l_4,l_1,l_2}\right)-\text{II}_{l_4,l_0,l_1}\right) 
\nonumber \\ &
+c_{10} (3 \log (2)
\text{II}_{l_0,l_4,l_1}-\frac{1}{2} \text{II}_{l_0,l_4,l_0,l_1}+\frac{3}{2}
\text{II}_{l_0,l_4,l_1,l_0}+3 \text{II}_{l_0,l_4,l_1,l_2})-2 c_{22} \text{II}_{l_1,l_2,l_1,l_1}\nonumber \\ & +2 c_4
\left(\text{II}_{l_1,l_1,l_1,l_0}+2 \text{II}_{l_1,l_1,l_1,l_2}\right)
+c_{26} \left(-18 \log (2)
\text{II}_{l_1,l_4,l_1}-\text{II}_{l_1,l_0,l_0,l_1}-2 \text{II}_{l_1,l_0,l_1,l_0}\right.\nonumber \\ &  -4
\text{II}_{l_1,l_0,l_1,l_2}+4 \text{II}_{l_1,l_0,l_2,l_1}+4 \text{II}_{l_1,l_1,l_0,l_0}+8
\text{II}_{l_1,l_1,l_0,l_2}+4 \text{II}_{l_1,l_1,l_1,l_1}+8 \text{II}_{l_1,l_1,l_2,l_0}\nonumber \\ &  +16
\text{II}_{l_1,l_1,l_2,l_2}+4 \text{II}_{l_1,l_2,l_0,l_1}-4 \text{II}_{l_1,l_2,l_1,l_0}-8
\text{II}_{l_1,l_2,l_1,l_2}-8 \text{II}_{l_1,l_2,l_2,l_1}+3 \text{II}_{l_1,l_4,l_0,l_1}\nonumber \\ & \left. -9
\text{II}_{l_1,l_4,l_1,l_0}-18 \text{II}_{l_1,l_4,l_1,l_2}\right)-c_{23}
\text{II}_{l_1,l_0,l_1,l_1}+2 c_{20}
\text{II}_{l_1,l_1,l_2,l_1}\Big\} 
\end{align}
and 
\begin{align}
\rho^{(3)}_{4m}(s) =& \frac{2\pi N^2}{9} \Big\{-\frac{32}{27} c_{24} f''\left(\bar{s}\right)+\frac{4}{27} c_{32}
f'\left(\bar{s}\right)-\frac{1}{9} c_{33} f\left(\bar{s}\right)+\frac{1}{3} c_9 \left(2
\text{II}_{r_2}-\text{II}_{l_1,\tilde{r}_3}\right)-\frac{2}{9} c_{28}
\text{II}_{\tilde{r}_3}\Big\}\nonumber \\ & 
+\frac{2\pi N}{9} \Big\{-48 c_1 f''\left(\bar{s}\right)+2 c_{29}
f'\left(\bar{s}\right)-\frac{1}{2} c_{31} f\left(\bar{s}\right)+2 c_7
\left(\text{II}_{r_1}-\text{II}_{l_0,\tilde{r}_3}\right) +3 c_{14} \text{II}_{\tilde{r}_3}
\nonumber \\ &
+c_{19} (\frac{3}{2}
\text{II}_{l_1,\tilde{r}_3}-3 \text{II}_{r_2})+c_8 \left(2
\text{II}_{l_2,\tilde{r}_3}-4 \text{II}_{r_3}\right)+c_{13} \left(2
\text{II}_{l_0,l_1,\tilde{r}_3}-4 \text{II}_{l_0,r_2}\right) +2 c_{16}
\text{II}_{r_0} \nonumber \\ & +c_{26} \left(-2
\text{II}_{l_1,l_1,\tilde{r}_3}-\text{II}_{l_1,r_0}+4 \text{II}_{l_1,r_2}\right)+2 c_{18}
\left(-\text{II}_{l_1,l_0,\tilde{r}_3}+\text{II}_{l_1,l_2,\tilde{r}_3}+\text{II}_{l_1,r_1}-
2 \text{II}_{l_1,r_3}\right)\Big\}
\end{align}
Here, we used notation for $f(\s)$ function and iterated integrals as described in Appendix \ref{notation}. The expressions for $c_i$ coefficients can be found in Appendix \ref{c-coeffs}. The iterated integrals with $l_i$ weights only can be straightforwardly written in terms of Goncharov's multiple polylogarithms, while those with elliptic kernels ($r_i$ or $\tilde{r}_i$ weights) in terms of one-fold integrals of harmonic polylogarithms and complete elliptic integrals as shown in \cite{3loop-masters}. The latter representations are already well suited for numerical evaluations. However, as we will see in the next section, much more faster numerics in the whole range of $s$-values can be  done with threshold and high energy expansions of spectral densities.

\section{Asymptotics and checks}
\begin{figure}
	\begin{center}
		\includegraphics[width=0.8\textwidth]{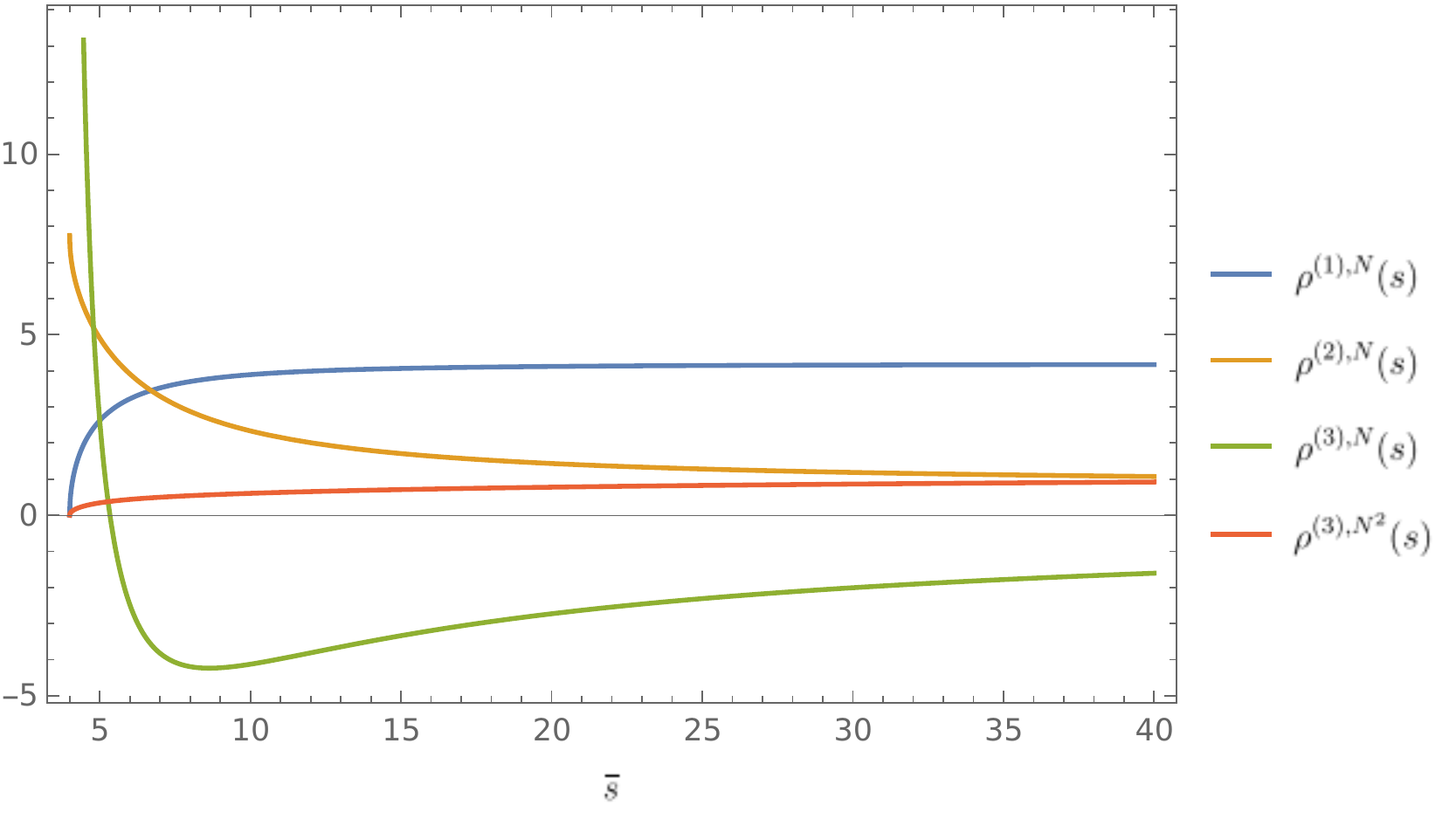}
		\caption{Values of one, two and three-loop spectral densities. Here $\rho^{(i)} = \rho^{(i),N} N +  \rho^{(i),N^2} N^2$}
	    \label{fig:densities}
	\end{center}
\end{figure}
\begin{figure}
	\begin{center}
		\includegraphics[width=0.8\textwidth]{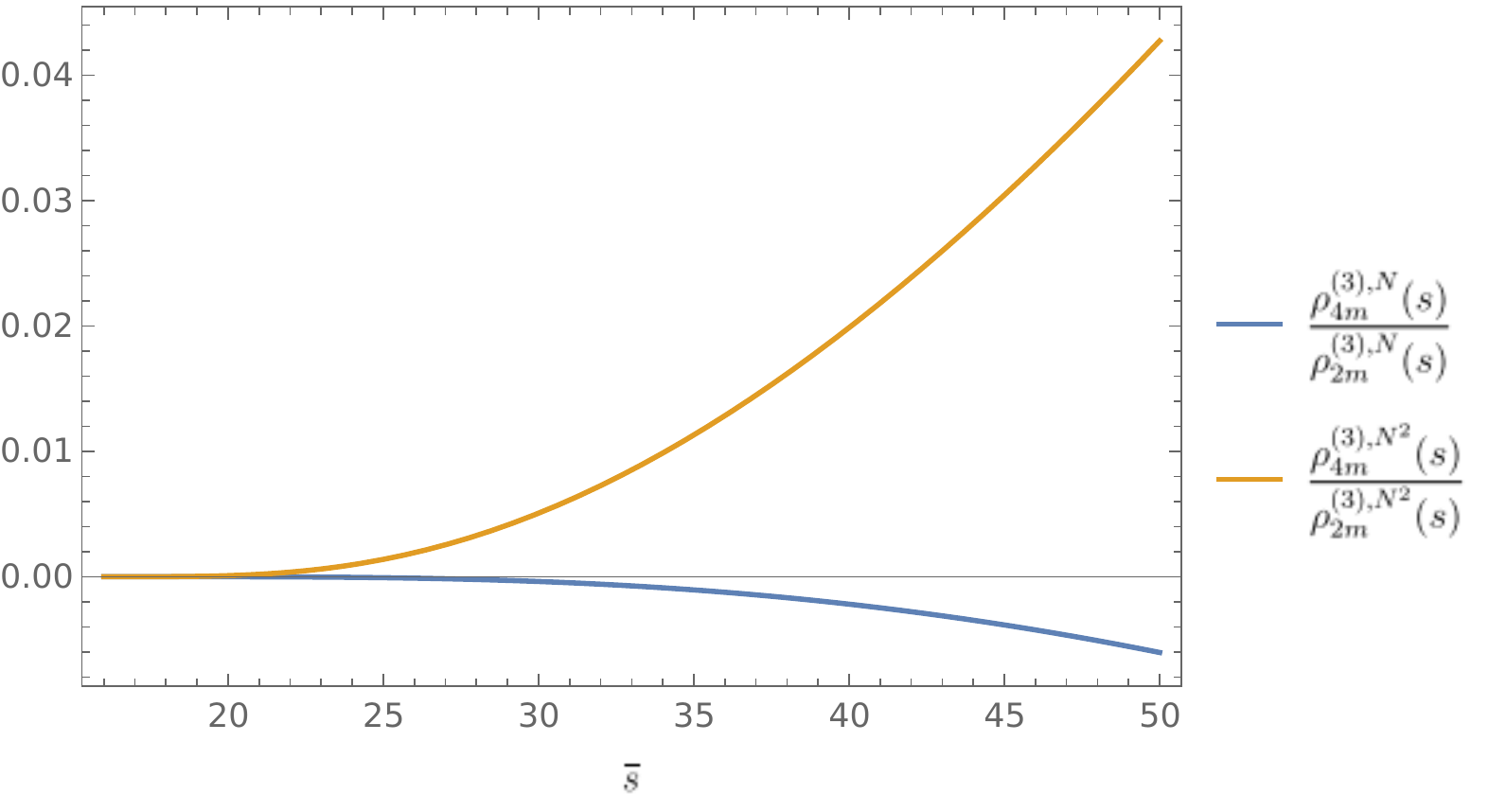}
		\caption{Ratio of $4m$ and $2m$ cuts contributions to three-loop spectral density. Here $\rho_*^{(i)} = \rho_*^{(i),N} N +  \rho_*^{(i),N^2} N^2$}
		\label{fig:ratio4mto2m}
	\end{center}
\end{figure}

The asymptotic expansions of obtained spectral densities can be done through the asymptotic expansions of iterated integrals as described in \cite{3loop-masters}. While it is straightforward to do asymptotic expansions in the threshold cases, the high-energy expansions are more involved and may require finding PLSQ relations \cite{PSLQ} for polylogarithmic constants at unity argument. To avoid this one can obtain required asymptotic expansions for master integrals themselves through the Frobenius solution of corresponding differential equations. The calculation of spectral densities with the asymptotic expressions for cut master integrals then gives  
\begin{align}
\rho^{(1)}_{thr} (s) =& ~\frac{2}{3} N\pi\beta (3-\beta^2)\, , \\
\rho^{(1)}_{high} (s) =& ~\frac{4N\pi}{3} \Big\{
1-\frac{6}{\s^2}-\frac{8}{\s^3}-\frac{18}{\s^4}
\Big\} + \OO \left(\frac{1}{\s^5}\right)\, , \\
\rho^{(2)}_{thr}(s) =& ~4N\pi\Big\{
\pi^2 - 8\beta +\frac{2\pi^2\beta^2}{3}-\frac{\pi^2\beta^4}{3}+\frac{4}{9}\beta^3 (-37+36\log 2+24\log\beta)
\Big\} + \OO\left(\beta^5\right)\, , \\
\rho^{(2)}_{high}(s) =& ~4N\pi \Big\{
1+\frac{12}{\s}+\frac{2(5+12\log\s)}{\s^2}  \nonumber\\ &  + \frac{16(-47+87\log\s)}{27\s^3}
+ \frac{-983+1218\log\s}{9\s^4}
\Big\} + \OO \left(\frac{1}{\s^5}\right)\, , \\
\rho^{(3)}_{2m, thr}(s) =& ~\frac{32N^2\pi}{9}\Big\{
4 (11-\pi^2)\beta - \frac{1}{9}(245-24\pi^2)\beta^3
\Big\} \nonumber\\ &  + \frac{8N\pi}{9}\Big\{
\frac{3\pi^4}{\beta} - 72\pi^2 + (351-70\pi^2+5\pi^4+48\pi^2\log 2 - 24\pi^2\log\beta -36\zeta_3)\beta  \nonumber \\
& + 2 (-43\pi^2 + 24\pi^2\log 2 + 48\pi^2\log\beta)\beta^2 \nonumber \\ &
+ (1411+51\pi^2 + \pi^4 - 1152\log 2- 42\pi^2\log 2 - (768 - 20\pi^2)\log\beta + 117\zeta_3)\beta^3 \nonumber \\ &
+ \frac{8}{75} (-947\pi^2 + 480\pi^2\log 2 + 960\pi^2\log\beta)\beta^4
\Big\} + \OO\left(\beta^5\right)\, , \\
\rho^{(3)}_{4m, thr}(s) =& ~\frac{\pi^2 N^2 (\s-16)^{9/2}}{516096}\Big\{
1 - \frac{629 (\s-16)}{2640} + \frac{10243 (\s-16)^2}{274560}
-\frac{7973 (\s-16)^3}{1647360}\Big\}\nonumber\\ &
+ \frac{\pi^2 N (\s-16)^{9/2}}{5160960}\Big\{
1 - \frac{7(\s-16)}{48} + \frac{307 (\s-16)^2}{27456} - \frac{193 (\s-16)^3}{658944} 
\Big\} + \OO \left(\s-16\right)^{17/2}\, , \\
\rho^{(3)}_{2m, high}(s) =& ~\frac{16N^2\pi}{81}
\Big\{
766 - 66\pi^2 - 265\log\s + 12\pi^2\log\s + 57\log^2\s - 6\log^3\s \nonumber \\
& + \frac{2}{\s}(-65+24\pi^2-216\log\s + 108\log^2\s) \Big\}  
+ \frac{2N\pi}{135}\Big\{
16065 - 900\pi^2 + 76\pi^4 \nonumber \\ &
 - 1440\pi^2\log 2 - 1440\zeta_3 + \log\s (-2340+360\pi^2-1440\zeta_3) + \frac{1}{\s} (-35100 + 3600\pi^2 \nonumber \\ & - 114\pi^4 - 11520\zeta_3 + \log\s (-7200+240\pi^2+1440\zeta^3) + (-7200+60\pi^2)\log^2\s \nonumber \\
& + 240\log^3\s - 30\log^4\s) 
\Big\} + \OO \left(\frac{1}{\s^2}\right)\, , \\
\rho^{(3)}_{4m,high}(s) =& ~\frac{8N^2\pi}{81}\Big\{
-1829 + 132\pi^2 + 216\zeta_3 + (584-24\pi^2)\log\s
- 114\log^2\s + 12\log^3\s \nonumber \\
& + \frac{1}{\s} (-1144-96\pi^2+1512\log\s - 432\log^2\s)
\Big\} + \frac{4N\pi}{135}\Big\{
-8100+450\pi^2\nonumber \\ & 
-38\pi^4+720\pi^2\log 2 + 720\zeta_3 +  (1170-180\pi^2+720\zeta_3)\log\s + \frac{1}{\s}(
18360 \nonumber \\ &
-1800\pi^2 + 57\pi^4 + 5760\zeta_3
- (6120+120\pi^2+720\zeta_3)\log\s \nonumber \\ &
+ (3600-30\pi^2)\log^2\s - 120\log^3\s + 15\log^4\s
)
\Big\} + \OO \left(\frac{1}{\s^2}\right)\, .
\end{align}
and results with more terms in the expansions can be found in accompanying Mathematica notebook. Using the latter it is easy to get very accurate representation of above spectral densities for the whole range of $\bar s$ values. The corresponding plot of different contributions can be found in Fig. \ref{fig:densities}. It is also interesting to compare ratio of $4m$ and $2m$ cuts contributions to three loop spectral density in the region of $\bar s$ where mass effects are important. The corresponding plot can be found in Fig. \ref{fig:ratio4mto2m}. From the latter we may conclude that the account of second threshold $\bar s = 16$ missed in the reconstruction of photon polarization operator performed in \cite{Baikov:1995ui} is not actually important.   

To check the results of our calculation we first performed it in an arbitrary gauge and made sure that the gauge dependence of spectral densities cancels. Second, we have numerically checked that the three moments\footnote{See Apendix \ref{moments}  for their definition.} $M_n$ of polarization operator computed in \cite{Baikov:1995ui} agree with a very high precision (at least $15$ digits) with  moments obtained with the use of our spectral densities and dispersion relation
\begin{equation}
\Pi (s) = \frac{1}{\pi}\int_0^{\infty}\frac{\rho (s')}{s' - s} \label{eq:dispersion}.
\end{equation} 
Moreover, we performed comparison with additionally calculated $97$ moments, see Appendix \ref{moments} and accompanying Mathematica notebook. This latter calculation is similar to the one we did for spectral densities except that we used uncut master integrals. The Frobenius solutions for the latter can be easily obtained from corresponding differential equations using standard techniques. Next, with the computed spectral densities we can deduce approximate analytical expressions for moments with large $n$ values. Indeed, it is easy to see making variable change in dispersion relation from $s'$ to $\beta' = \sqrt{1-4m^2/s'}$ that
\begin{equation}
M_n^{(i)} \simeq \int_0^1 d\beta' \frac{2\beta'}{\pi} (1-\beta'^2)^{n-1} \rho^{(i)}_{thr}(\beta')\, ,
\end{equation}
where for large $n$ values the largest contribution to integral comes from small $\beta'$ values and for three-loop spectral density it is sufficient to consider only $2m$ cut contribution. At one loop due to finite $\beta'$ expansion of spectral density this expression is actually exact and we have 
\begin{equation}
M_n^{(1)} = \frac{(1+n)\sqrt{\pi}\Gamma (n)}{\Gamma (\frac{5}{2}+n)} N\, .\label{eq:moment1loop}
\end{equation}    
At two and three loops the formulae are approximate and are given by
\begin{align}
M_n^{(2)}\simeq &\sum_{i=0}^{\text{order}_\beta}\frac{\Gamma (\frac{1+i}{2})\Gamma (n)}{4\Gamma (\frac{1+i+2n}{2})}\Big\{
2 d_{1,i}^{(2)} + d_{2,i}^{(2)}
\Big[
\Psi^{(0)} \Big(\frac{1+i}{2}\Big) - \Psi^{(0)}\Big(
\frac{1+i+2n}{2}
\Big)
\Big]
\Big\}\, , \label{eqn:moment2loop-largen} \\
M_n^{(3)}\simeq  &\sum_{i=0}^{\text{order}_\beta}\frac{\Gamma (\frac{1+i}{2})\Gamma (n)}{8 \Gamma (\frac{1+i+2n}{2})} \Big\{
4 d_{1,i}^{(3)} + 2 d_{2,i}^{(3)}
\left[
\Psi^{(0)} \left(\frac{1+i}{2}\right) - \Psi^{(0)}\left(
\frac{1+i+2n}{2}
\right)
\right]
 \nonumber \\
&+ d_{3,i}^{(3)}
\Big[
\Big(
\Psi^{(0)}\Big(
\frac{1+i}{2}
\Big) - \Psi^{(0)}\Big(
\frac{1+i+2n}{2}
\Big)
\Big)^2 + \Psi^{(1)}\Big(\frac{1+i}{2}\Big) - \Psi^{(1)}\Big(
\frac{1+i+2n}{2}
\Big)
\Big]
\Big\}\, ,  \label{eqn:moment3loop-largen}
\end{align}
where  $d$-coefficients are defined as 
\begin{align}
\frac{2\beta}{\pi}\rho^{(2)}_{thr}(\beta)\simeq &\sum_{i=0}^{\text{order}_\beta} d_{1,i}^{(2)}\beta^i + d_{2,i}^{(2)}\beta^i\log\beta\, , \\
\frac{2\beta}{\pi}\rho^{(3)}_{thr}(\beta)\simeq &\sum_{i=0}^{\text{order}_\beta} d_{1,i}^{(3)}\beta^i   
+ d_{2,i}^{(3)}\beta^i\log\beta
+ d_{3,i}^{(3)}\beta^i\log^2\beta
\, .
\end{align}
Here $\text{order}_\beta$ is the expansion order in $\beta$ of corresponding spectral densities and 
\begin{equation}
\Psi^{(\nu)}(z) = \frac{d^{\nu + 1}\log \Gamma (z)}{dz^{n+1}}
\end{equation}
is the usual polygamma function. With $\text{order}_\beta = 80$ the approximate expression for $M_1^{(2)}$ is accurate with $0.04$ percent level, $M_{10}^{(2)}$ with $10^{-10}$ percent level and $M_{50}^{(2)}$ with $10^{-26}$ percent level. For $\text{order}_\beta = 10$ the corresponding errors are $2$, $10^{-3}$ and $10^{-6}$ percents. With  $\text{order}_\beta = 120$ the accuracy for approximate expression of $M_1^{(3)}$ is $0.4$ percent, for $M_{10}^{(3)}$ it is $10^{-10}$ percent and for $M_{50}^{(3)}$ - $10^{-32}$ percents. For lower value of $\text{order}_\beta = 10$ the corresponding values are $13$, $10^{-3}$ and $10^{-7}$ percents. So, we see that for sufficiently large values of moments our approximate formula are very accurate  except for a few first moments.  

\section{Conclusion}

In the present paper we performed calculation of three-loop photon spectral density in the framework of QED with $N$ different electron flavors. The obtained results contain both exact and asymptotic expressions. The exact result were written in terms of iterated integrals, which reduce either to Goncharov's polylogarithms or to one-fold integrals of harmonic polylogarithms and complete elliptic integrals.  The asymptotic expressions on the other hand allow to have very accurate expressions for the spectral density in the whole $s$ range. It is shown that the obtained spectral density correctly reproduces calculated first hundred moments of the photon polarization operator. In addition we supply approximate analytical formula for the moments of photon polarization operator. The latter are very accurate for almost all moments except maybe first few.

\paragraph*{Acknowledgments} The author is grateful to R.N. Lee for interest to this work and valuable discussions. This work was supported by Russian Science Foundation, grant 20-12-00205.

\appendix

\section{On-shell renormalization constants}\label{renorm-constants}

The required on-shell renormalization constants for QED with $N$ similar electron species can be extracted from renormalization of 2-loop photon and electron self-energies. For photon wave function renormalization ($A_0 = Z_A A$) we have ($\alpha = \frac{e^2}{4\pi}$):
\begin{equation}
Z_A = 1+ Z_A^{(1)} \left(
\frac{\alpha \mu^{2\ep} e^{-\ep\gamma_E}}{(4\pi)^{1-\ep}m^{2\ep}}
\right)
+ Z_A^{(2)} \left(
\frac{\alpha \mu^{2\ep} e^{-\ep\gamma_E}}{(4\pi)^{1-\ep}m^{2\ep}}
\right)^2 + \ldots\, , \label{eq:wave-function-renormalization}
\end{equation}
where
\begin{align}
Z_A^{(1)} &= N\Big\{ 
-\frac{4}{3\ep} -\frac{\pi^2\ep}{9} + \frac{4}{9}\ep^2\zeta_3- \frac{\pi^4\ep^3}{120} + \frac{\ep^4}{135} (5\pi^2\zeta_3 + 36\zeta_5)
\Big\}+ \ldots \, , \\
Z_A^{(2)} &= N\Big\{
-\frac{2}{\ep} - 15 - \frac{1}{6}(93+2\pi^2)\ep + \ep^2 (-\frac{223}{4}-\frac{5\pi^2}{2}+\frac{4\zeta_3}{3})
\Big\} + \ldots 
\end{align}
Here $m$ is the electron pole mass, $\mu$ - dimensional parameter entering dimensional regularization and $e$ is on-shell electron charge at $q^2 = 0$. The charge renormalization constant ($\alpha_0 = Z_{\alpha}\alpha\mu^{2\ep}$) is then obtained using Ward identity as 
\begin{equation}
Z_{\alpha} = Z_A^{-1}\, . \label{eq:charge-renormalization}
\end{equation}
Similar from electron self-energy for on-shell mass renormalization constant ($m_0 = Z_m m$) we get
\begin{equation}
Z_m = 1+ Z_m^{(1)} \left(
\frac{\alpha \mu^{2\ep} e^{-\ep\gamma_E}}{(4\pi)^{1-\ep}m^{2\ep}}
\right)
+ Z_m^{(2)} \left(
\frac{\alpha \mu^{2\ep} e^{-\ep\gamma_E}}{(4\pi)^{1-\ep}m^{2\ep}}
\right)^2 + \ldots\, , \label{eq:mass-renormalization}
\end{equation}
where
\begin{align}
Z_m^{(1)} = &-\frac{3}{\ep} - 4 -(8+\frac{\pi^2}{4})\ep + \ep^2 (-16-\frac{\pi^2}{3}+\zeta_3) + \ep^3 (-32-\frac{2\pi^2}{3}-\frac{3\pi^4}{160}+\frac{4\zeta_3}{3}) \nonumber \\
&+\ep^4 (-64 - \frac{4\pi^2}{3} - \frac{\pi^4}{40} + \frac{8\zeta_3}{3} + \frac{1}{12}\pi^2\zeta_3 + \frac{3\zeta_5}{5}) + \ldots \, , \\
Z_m^{(2)} = &~\frac{9}{2\ep^2} + \frac{45}{4\ep} + \frac{199}{8} - \frac{17\pi^2}{4} + 8\pi^2\log 2 - 12\zeta_3 + \ep (\frac{677}{16} - \frac{205\pi^2}{8} + \frac{14\pi^4}{5}   \nonumber \\
&+ 48\pi^2\log 2 - 16\pi^2\log^2 2 - 8\log^4 2 - 192 a_4 - 135\zeta_3)
+ \ep^2 (-\frac{110857}{128}   \nonumber \\ &+ \frac{7039\pi^2}{192} + \frac{209\pi^4}{64} - 54\pi^2\log 2 + 8\pi^2\log^2 2 + 4\log^4 2
+ 96a_4 + \frac{4915}{24}\zeta_3  \nonumber \\
&+ \frac{443}{36}\pi^2\zeta_3 + \frac{575}{2}\zeta_5
) + N\Big\{
-\frac{2}{\ep^2} + \frac{5}{3\ep} + \frac{143}{6} - 3\pi^2 
+ \ep (\frac{1133}{12}-\frac{235\pi^2}{18} \nonumber \\ 
&+16\pi^2\log 2 -\frac{164}{3}\zeta_3) + \ep^2 (\frac{8135}{24} - \frac{1585\pi^2}{36}
+\frac{827\pi^4}{360} + 80\pi^2\log 2  \nonumber \\
&- 32\pi^2\log^2 2 - 16\log^4 2 - 384a_4 -\frac{2530}{9}\zeta_3 ) \Big\} + \ldots
\end{align}
and $a_4 = \text{Li}_4 \left(\frac{1}{2}\right) = \sum_{n=1}^{\infty}\frac{1}{2^n n^4}$. Note, that at $N=1$ these renormalization constants reduce to already known QED values \cite{QEDrenorm1,QEDrenorm2,QEDrenorm3}. 

\section{Notation for iterated integrals}\label{notation}
The iterated integrals present in current paper involve the following set of integration kernels or weights:
\begin{gather}\label{eq:weights1}
    l_0(\s) = \frac{1}{\s}, \quad l_1(\s) = \frac{1}{\s \beta}, \quad l_2(\s) = \frac{1}{\s -4}, 
    \quad l_4(\s) = \frac{1}{\s-1}\,,\\
    r_k(\s) = \frac{f(\s)}{\s \beta^k}\theta(\s-16)\quad (k=0,1,2,3),\\
    \tilde{r}_3(\s) =\frac{8 \beta  (\s+2) f(\s)}{(\s-16) (\s-4)^2}\theta(\s-16)\,,
\end{gather}
where 
\begin{gather}\label{eq:fdef}
    f(\s)=\frac{16 (\s-16)}{\s}\left[
    \mathrm{K}(1-k_{-})\mathrm{K}(k_{+}) - \mathrm{K}(k_{-})\mathrm{K}(1-k_{+})
    \right],\\
    k_{\pm}=\frac12\left[1\pm\left(1-\frac{8}{\s}\right) \sqrt{1-\frac{16}{\s}}+\frac{16}{\s} \sqrt{1-\frac{4}{\s}}\right]\,.
\end{gather}
and $K$ is complete elliptic integral of first kind. The required iterated integrals are then defined as
\begin{equation}
\II_{w_n,\ldots w_1} = I(w_n,\ldots w_1|\s)= \int\limits_{\s>s_n>\ldots >s_1>4} \prod_{k=1}^{n} ds_k w_k(s_k)\, ,
\end{equation}
where weights $w_k(s)$ take values from the set above.

\section{Expressions for $c_i$ coefficients}\label{c-coeffs}

For $c_i$ coefficients present in the expression for three-loop photon spectral density we have the following expressions 
\begin{align}
c_1 =& -\frac{29}{8 (\beta +1)}+\frac{15}{4 (\beta +1)^2}+\frac{29}{8 (\beta -1)}+\frac{15}{4 (\beta
	-1)^2}-\frac{1}{4}\, ,  \;
c_2 = \frac{3 \beta }{4}-\frac{\beta ^3}{4}\, ,\\ 
c_3 =& -\frac{\beta ^4}{4}+\frac{\beta ^2}{2}-\frac{1}{4}\, ,\; 
c_4 = -\frac{3 \beta ^4}{4}+\frac{7 \beta ^2}{2}-\frac{11}{4}\, , \;
c_5 = -\frac{69 \beta ^3}{4}+\frac{113 \beta }{4}+\frac{3}{\beta }\, , \\
c_6 =& -\frac{15 \beta ^3}{2}+\frac{35 \beta }{4}+\frac{3}{4 \beta }\, , \;
c_7 = -\frac{15 \beta ^3}{4}+8 \beta +\frac{3}{4 \beta }\, , \;
c_8 = -\frac{9 \beta ^3}{4}+\frac{7 \beta }{2}+\frac{3}{4 \beta }\, , \\
c_9 =& -\frac{5 \beta ^4}{4}+\frac{5 \beta ^2}{2}+\frac{3}{4}\, , \;
c_{10} = -\frac{3 \beta ^4}{4}+\frac{\beta ^2}{2}+\frac{17}{4}\, , \;
c_{11} = -\frac{\beta ^4}{2}+\frac{11 \beta ^2}{4}+\frac{23}{4}\, , \\
c_{12} =& -\frac{\beta ^4}{4}+\frac{\beta ^2}{2}+\frac{3}{4}\, , \; c_{13} = -\frac{\beta ^4}{4}+\frac{\beta ^2}{2}+\frac{7}{4}\, , \; 
c_{14} = -\frac{\beta ^3}{4}+\frac{3 \beta }{4}+\frac{2}{\beta }\, , \\
c_{15} = & -\frac{\beta ^4}{4}+\beta ^2+\frac{5}{4}\, , \;
c_{16} = \frac{9}{4}-\frac{\beta ^4}{4}\, ,\; c_{17} = \frac{5}{4}-\frac{\beta ^4}{4}\, , \; c_{18} = -\frac{\beta ^4}{4}+\frac{3 \beta ^2}{4}-\frac{3}{2}\, , \\
c_{19} =& -\frac{5 \beta ^4}{4}+\frac{9 \beta ^2}{2}-\frac{5}{4}\, , \; c_{20} = -\frac{7 \beta ^4}{4}+\frac{13 \beta ^2}{2}-\frac{35}{4}\, , \; c_{21} = -\frac{9 \beta ^4}{4}+8 \beta ^2-\frac{47}{4}\, , \\
c_{22} =& -\frac{11 \beta ^4}{4}+12 \beta ^2-\frac{45}{4}\, , \; c_{23} = -\frac{17 \beta ^4}{4}+\frac{33 \beta ^2}{2}-\frac{81}{4}\, , \\ c_{24} =& -\frac{407}{16 (\beta +1)}+\frac{1829}{16 (\beta +1)^2}+\frac{407}{16 (\beta
	-1)}+\frac{1829}{16 (\beta -1)^2}-\frac{771}{4}\, , \\
c_{25} =& -\frac{1}{4} \left(15 \pi ^2-236\right) \beta ^3+\frac{3}{4} \left(27 \pi ^2-340\right) \beta
+\frac{9}{2 \beta }\, ,  \; c_{26} = -\frac{\beta ^5}{4}+\frac{\beta ^3}{4}+\frac{5 \beta }{4}+\frac{3}{4 \beta }\, , \\
c_{27} =& -\frac{15 \beta ^3}{4}-\frac{3}{4 \beta ^3}+4 \beta +\frac{10}{\beta }\, , \; c_{28} = -\frac{39 \beta ^3}{4}-\frac{3}{4 \beta ^3}-\frac{19 \beta }{2}-\frac{7}{2 \beta }\, , \\
c_{29} =& -\frac{39 \beta ^2}{2}+\frac{6507}{16 \left(4 \beta ^2-3\right)}+\frac{77}{2 (\beta
	-1)}-\frac{77}{2 (\beta +1)}+\frac{1997}{16}\, , \\
c_{30} =& -\frac{1}{4} \left(15 \pi ^2-236\right) \beta ^4+\frac{1}{2} \left(15 \pi ^2-167\right) \beta
^2+\frac{9}{\beta ^2}+\frac{9}{4} \left(\pi ^2-52\right)\, , \\
c_{31} =& -2 \beta ^4-\frac{5 \beta ^2}{8}-\frac{567}{4 \left(4 \beta ^2-3\right)}+\frac{6507}{64
	\left(4 \beta ^2-3\right)^2}+\frac{24}{\beta ^2}+\frac{1197}{64}\, , \\
c_{32} =& -\frac{967 \beta ^2}{2}+\frac{72063}{16 \left(4 \beta ^2-3\right)}+\frac{18}{\beta
	^2}+\frac{1245}{4 (\beta -1)}-\frac{1245}{4 (\beta +1)}+\frac{21117}{16}\, , \\
c_{33} =& -4 \beta ^4+\frac{3}{\beta ^4}+\frac{481 \beta ^2}{8}-\frac{1693}{4 \left(4 \beta
	^2-3\right)}+\frac{24021}{64 \left(4 \beta ^2-3\right)^2}+\frac{11}{\beta
	^2}+\frac{1011}{64}\, , \\
c_{34} =& -\frac{1}{2} \beta ^5 \left(-6+3 \pi ^2-4 \log ^2(2)\right)-\frac{7 \pi ^2 \beta
	^4}{4}+\frac{1}{4} \beta ^3 \left(191+6 \pi ^2-8 \log ^2(2)+24 \log (2)\right)\nonumber \\ &+\frac{13 \pi
	^2 \beta ^2}{2}+\frac{\beta}{2}   \left(-306+15 \pi ^2-20 \log ^2(2)+20 \log
(2)\right)+\frac{3 \left(1+6 \pi ^2-8 \log ^2(2)\right)}{4 \beta }-\frac{35 \pi ^2}{4}\, , \\
c_{35} =& -2 \pi ^2 \beta ^4+\frac{1}{4} \beta ^3 \left(9+14 \pi ^2+32 \log (2)\right)+4 \pi ^2 \beta
^2\nonumber \\ &+\frac{24 \beta }{\beta ^2+3}-\frac{32 \beta }{\left(\beta ^2+3\right)^2}-\frac{1}{4}
\beta  \left(-9+14 \pi ^2+96 \log (2)\right)+6 \pi ^2\, , \\
c_{36} =& -2 \pi ^2 \beta ^4-\frac{1}{4} \beta ^3 \left(9+66 \pi ^2+32 \log (2)\right)-\frac{24 \beta
}{\beta ^2+3}+\frac{32 \beta }{\left(\beta ^2+3\right)^2}\nonumber \\ & +\frac{3 \pi ^2}{\beta
}+\frac{1}{4} \beta  \left(-9+118 \pi ^2+96 \log (2)\right)+10 \pi ^2\, , \\
c_{37} =& -\frac{1}{4} \beta ^4 \left(-213+4 \pi ^2+48 \log (2)\right)-\frac{132}{\beta
	^2+3}+\frac{432}{\left(\beta ^2+3\right)^2}\nonumber \\ & -\frac{384}{\left(\beta ^2+3\right)^3}-2 \beta
^2 \left(39+\pi ^2-12 \log (2)\right)+\frac{3}{4} \left(-295+4 \pi ^2-16 \log (2)\right)\, , \\
c_{38} =& -2 \pi ^2 \beta ^5-\frac{1}{4} \beta ^4 \left(-41+6 \pi ^2-64 \log (2)\right)+2 \pi ^2 \beta
^3+\frac{264}{\beta ^2+3} -\frac{864}{\left(\beta ^2+3\right)^2}+\frac{768}{\left(\beta
	^2+3\right)^3} \nonumber \\ & +\frac{1}{2} \beta ^2 \left(3+14 \pi ^2-64 \log (2)\right)+10 \pi ^2 \beta
+\frac{6 \pi ^2}{\beta }+\frac{1}{4} \left(-247-22 \pi ^2-192 \log (2)\right)\, , \\
c_{39} =& -2 \pi ^2 \beta ^5+\frac{1}{4} \beta ^4 \left(-41+14 \pi ^2-64 \log (2)\right)+2 \pi ^2 \beta
^3-\frac{264}{\beta ^2+3}+\frac{864}{\left(\beta ^2+3\right)^2}-\frac{768}{\left(\beta
	^2+3\right)^3}\nonumber \\ & -\frac{1}{2} \beta ^2 \left(3+26 \pi ^2-64 \log (2)\right)+10 \pi ^2 \beta
+\frac{6 \pi ^2}{\beta }+\frac{1}{4} \left(247+70 \pi ^2+192 \log (2)\right)\, , \\
c_{40} =& -\frac{1}{2} \beta ^5 \log (2)+\frac{1}{4} \beta ^3 (2 \log (2)-3)+\frac{5}{4} \beta  (2 \log
(2)-1)+\frac{3 \log (2)}{2 \beta }\, , \\
c_{41} =& -\frac{1}{2} \beta ^5 \log (2)+\frac{1}{4} \beta ^3 (33+2 \log (2))+\frac{1}{4} \beta  (10
\log (2)-59)+\frac{6 \log (2)-6}{4 \beta }\, , \\ 
c_{42} =& -\frac{1}{2} \beta ^5 \log (2)+\frac{1}{4} \beta ^3 (48+2 \log (2))+\frac{1}{4} \beta  (10
\log (2)-91)+\frac{6 \log (2)-9}{4 \beta }\, , \\ 
c_{43} =& -\frac{1}{2} \beta ^4 (2 \log (2)-19)+\frac{1}{4} \beta ^2 (8 \log (2)-101)-3+3 \log (2)\, , \\
c_{44} =& -\frac{1}{4} \beta ^4 (31+12 \log (2))+\frac{7}{2} \beta ^2 (1+4 \log (2))-\frac{11}{4} (5+4
\log (2))\, , \\
c_{45} =& -2 \pi ^2 \left(\pi ^2-3\right) \beta ^5-12 \pi ^2 \beta ^4 (19+2 \log (2))+6 \pi ^2 \beta ^2
(101+8 \log (2))\nonumber \\ & -\frac{288 \beta  \log (2)}{\beta ^2+3} +\frac{384 \beta  \log
	(2)}{\left(\beta ^2+3\right)^2}+\frac{6 \pi ^4}{\beta }+\frac{1}{4} \beta ^3 \left(-1764
\zeta (3)-615+628 \pi ^2+8 \pi ^4\right.\nonumber \\ & \left. -192 \log ^2(2)-108 \log (2)-840 \pi ^2 \log
(2)\right)+\frac{1}{4} \beta  \left(1764 \zeta (3)+2757-1872 \pi ^2\right. \nonumber\\ & \left. +40 \pi ^4+576 \log
^2(2) -108 \log (2)+1416 \pi ^2 \log (2)\right)+72 \pi ^2 (1+\log (2))\, , \\
c_{46} =& -24 \pi ^2 \beta ^5 \log (2)+12 \pi ^2 \beta ^3 (2 \log (2)-33)-\frac{96 (33 \log
(2)-4)}{\beta ^2+3}+\frac{384 (27 \log (2)-1)}{\left(\beta ^2+3\right)^2}\nonumber \\ & -\frac{9216 \log
(2)}{\left(\beta ^2+3\right)^3}+12 \pi ^2 \beta  (59+10 \log (2))+\frac{72 \pi ^2 (1+\log
(2))}{\beta }+\frac{1}{4} \beta ^4 \left(756 \zeta (3)+597\right. \nonumber\\ & \left. +104 \pi ^2-384 \log ^2(2)-492
\log (2)+168 \pi ^2 \log (2)\right)-\frac{1}{2} \beta ^2 \left(1692 \zeta (3)+621+152 \pi
^2\right. \nonumber\\ & \left. -384 \log ^2(2)+36 \log (2)+312 \pi ^2 \log (2)\right)+\frac{1}{4} \left(2052 \zeta
(3)-243-1144 \pi ^2\right.\nonumber\\ & \left. +1152 \log ^2(2)+2964 \log (2)+840 \pi ^2 \log (2)\right)\, .
\end{align}

\section{Moments of photon correlation function}\label{moments}

The moments of photon correlation function are given by:
\begin{equation}
\Pi (\s) = \sum_{n>0} M_n \left(\frac{\s}{4}\right)^n\, ,
\end{equation}
where
\begin{equation}
M_n = M^{(1)}_n \left(\frac{\alpha}{4\pi}\right) 
+ M^{(2)}_n \left(\frac{\alpha}{4\pi}\right)^2
+ M^{(3)}_n \left(\frac{\alpha}{4\pi}\right)^3 + \ldots
\end{equation}
and the first five moments at one, two and three loops are given by 
\begin{align}
M^{(1)}_1 =& ~\frac{16 N}{15}\, ,  & M^{(2)}_1 =& ~\frac{1312 N}{81}\, ,\\
M^{(1)}_2 =& ~\frac{16 N}{35}\, ,  & M^{(2)}_2 =& ~\frac{7184 N}{675}\, ,\\
M^{(1)}_3 =& ~\frac{256 N}{945}\, , & M^{(2)}_3 =& ~\frac{3998656 N}{496125}\, ,  \\
M^{(1)}_4 =& ~\frac{128 N}{693}\, , & M^{(2)}_4 =& ~\frac{831776 N}{127575}\, ,\\
M^{(1)}_5 =& ~\frac{2048 N}{15015}\, , & M^{(2)}_5 =& ~\frac{6918163456 N}{1260653625}\, .
\end{align}
\begin{align}
M^{(3)}_1 =& ~N^2 \left(\frac{406 \zeta_3}{27}-\frac{45628}{729}+\frac{256 \pi ^2}{45}\right)+N
\left(\frac{22781 \zeta_3}{108}-\frac{8687}{54}+\frac{32 \pi ^2}{3}-\frac{256}{15} \pi ^2
\log (2)\right)\, , \\
M^{(3)}_2 =& ~N^2 \left(\frac{14203 \zeta_3}{1152}-\frac{1520789}{25920}+\frac{512 \pi ^2}{105}\right)\nonumber \\ & +N
\left(\frac{4857587 \zeta_3}{2880}-\frac{223404289}{116640}+\frac{64 \pi
	^2}{7}-\frac{512}{35} \pi ^2 \log (2)\right)\, , \\
M^{(3)}_3 =& ~N^2 \left(\frac{12355 \zeta_3}{864}-\frac{83936527}{1458000}+\frac{4096 \pi
	^2}{945}\right)\nonumber \\ &+N \left(\frac{33067024499 \zeta_3}{3225600}-\frac{885937890461}{72576000}+\frac{512 \pi ^2}{63}-\frac{4096}{315} \pi ^2
\log (2)\right)\, , \\
M^{(3)}_4 =& ~N^2 \left(\frac{2522821 \zeta_3}{147456}-\frac{129586264289}{2239488000}+\frac{8192 \pi^2}{2079}\right)\nonumber \\ & +N \left(\frac{1507351507033 \zeta_3}{25804800}-\frac{269240669884818833}{3840721920000}+\frac{5120 \pi^2}{693}-\frac{8192}{693} \pi ^2 \log (2)\right)\, , \\
M^{(3)}_5 =& ~N^2 \left(\frac{1239683 \zeta_3}{61440}-\frac{512847330943}{8692992000}+\frac{32768 \pi^2}{9009}\right)\nonumber \\ & +N \left(\frac{939939943788973 \zeta_3}{2980454400}-\frac{360248170450504167133}{950578675200000}+\frac{20480 \pi^2}{3003}-\frac{32768 \pi ^2 \log (2)}{3003}\right)\, .
\end{align}
The rest of moments up to $n=100$ can be found in accompanying Mathematica notebook.

\bibliographystyle{JHEP}
\bibliography{PO}

\providecommand{\href}[2]{#2}\begingroup\raggedright\begin{thebibliography}{10}

\bibitem{berestetskii1982quantum}
V.~B. Berestetskii, E.~M. Lifshitz, and L.~P. Pitaevskii, {\em Quantum
  Electrodynamics: Volume 4}, vol.~4.
\newblock Butterworth-Heinemann, 1982.

\bibitem{kallen1955k}
G.~K{\"a}ll{\'e}n and A.~Sabry, {\it Fourth order vacuum polarization},  {\em
  Dan. Mat Fys. Medd.} {\bf 29} (1955), no.~17 556.

\bibitem{Baikov:1995ui}
P.~Baikov and D.~J. Broadhurst, {\it {Three loop QED vacuum polarization and
  the four loop muon anomalous magnetic moment}},  in {\em {4th International
  Workshop on Software Engineering and Artificial Intelligence for High-energy
  and Nuclear Physics}}, pp.~0167--172, 4, 1995.
\newblock \href{http://arxiv.org/abs/hep-ph/9504398}{{\tt hep-ph/9504398}}.

\bibitem{kinoshita1999sixth}
T.~Kinoshita and M.~Nio, {\it Sixth-order vacuum-polarization contribution to
  the lamb shift of muonic hydrogen},  {\em Physical review letters} {\bf 82}
  (1999), no.~16 3240.

\bibitem{kinoshita1999accuracy}
T.~Kinoshita and M.~Nio, {\it Accuracy of calculations involving $\alpha$ 3
  vacuum-polarization diagrams: Muonic hydrogen lamb shift and muon g- 2},
  {\em Physical Review D} {\bf 60} (1999), no.~5 053008.

\bibitem{kinoshita1983parametric}
T.~Kinoshita and W.~Lindquist, {\it Parametric formula for the sixth-order
  vacuum polarization contribution in quantum electrodynamics},  {\em Physical
  Review D} {\bf 27} (1983), no.~4 853.

\bibitem{FeynArts}
T.~Hahn, {\it {Generating Feynman diagrams and amplitudes with FeynArts 3}},
  {\em Comput. Phys. Commun.} {\bf 140} (2001) 418--431,
  [\href{http://arxiv.org/abs/hep-ph/0012260}{{\tt hep-ph/0012260}}].

\bibitem{ChetTka1981}
K.~G. Chetyrkin and F.~V. Tkachov, {\it Integration by parts: {T}he algorithm
  to calculate $\beta$-functions in 4 loops},  {\em Nucl.~Phys.~B} {\bf 192}
  (1981) 159.

\bibitem{Tkachov1981}
F.~V. Tkachov, {\it A theorem on analytical calculability of 4-loop
  renormalization group functions},  {\em Physics Letters B} {\bf 100} (Mar.,
  1981) 65--68.

\bibitem{3loop-masters}
R.~N. Lee and A.~I. Onishchenko, {\it {Master integrals for bipartite cuts of
  three-loop photon self energy}},  {\em JHEP} {\bf 04} (2021) 177,
  [\href{http://arxiv.org/abs/2012.04230}{{\tt arXiv:2012.04230}}].

\bibitem{Lee:2019wwn}
R.~Lee and A.~Onishchenko, {\it {$\epsilon$-regular basis for
  non-polylogarithmic multiloop integrals and total cross section of the
  process $e^+e^-\to 2(Q\bar Q)$}},  {\em JHEP} {\bf 12} (2019) 084,
  [\href{http://arxiv.org/abs/1909.07710}{{\tt arXiv:1909.07710}}].

\bibitem{PSLQ}
H.~Ferguson and D.~Bailey, {\it A polynomial time, numerically stable integer
  relation algorithm}, .

\bibitem{QEDrenorm1}
N.~Gray, D.~J. Broadhurst, W.~Grafe, and K.~Schilcher, {\it {Three Loop
  Relation of Quark (Modified) Ms and Pole Masses}},  {\em Z. Phys. C} {\bf 48}
  (1990) 673--680.

\bibitem{QEDrenorm2}
D.~J. Broadhurst, N.~Gray, and K.~Schilcher, {\it {Gauge invariant on-shell
  Z(2) in QED, QCD and the effective field theory of a static quark}},  {\em Z.
  Phys. C} {\bf 52} (1991) 111--122.

\bibitem{QEDrenorm3}
K.~Melnikov and T.~van Ritbergen, {\it {The Three loop on-shell renormalization
  of QCD and QED}},  {\em Nucl. Phys. B} {\bf 591} (2000) 515--546,
  [\href{http://arxiv.org/abs/hep-ph/0005131}{{\tt hep-ph/0005131}}].

\end{thebibliography}\endgroup

\end{document}